\documentstyle[12pt]{article}

\renewcommand{\thefootnote}{\fnsymbol{footnote}}
\newcommand{\vs}[1]{\vspace{#1 mm}}

\makeatletter
\def\eqnarray{%
 \stepcounter{equation}%
 \let\@currentlabel=\theequation
 \global\@eqnswtrue
 \global\@eqcnt\z@
 \tabskip\@centering
 \let\\=\@eqncr
 $$\halign to \displaywidth\bgroup\@eqnsel\hskip\@centering
 $\displaystyle\tabskip\z@{##}$&\global\@eqcnt\@ne
 \hfil$\displaystyle{{}##{}}$\hfil
 &\global\@eqcnt\tw@$\displaystyle\tabskip\z@{##}$\hfil
 \tabskip\@centering&\llap{##}\tabskip\z@\cr}
\makeatother

\begin{document}
\baselineskip 5mm

\newcommand{\ov}{\overline}
\newcommand{\ep}{\epsilon}

\renewcommand{\theequation}{\thesection.\arabic{equation}}

\newcommand{\apsect}[1]{\setcounter{equation}{0}
                      \setcounter{subsection}{0}
                      \addtocounter{section}{1}
                      \section*{Appendix\, #1}}

\begin{titlepage}
\setcounter{page}{0}
\begin{flushright}
KEK-TH 858\\
December 2002\\
\end{flushright}

\vs{6}
\begin{center}
{\LARGE  The Fuzzy $S^4$ by Quantum Deformation}
\vspace{2cm}

\vs{6}
{\large
Shogo Aoyama$^{1}$\footnote{e-mail: spsaoya@ipc.shizuoka.ac.jp}\ \ \ \ and 
\ \ Takahiro Masuda$^{2}$\footnote{e-mail: stmasud@post.kek.jp} 
}\\
\vs{6}
{\em ${}^1$ Department of Physics, Shizuoka University \\
          Ohya 836, Shizuoka, Japan  \\
 \   \\         
      ${}^2$ High Energy Accelerator Research Organization (KEK),\\ 
               Tsukuba, Ibaraki 305-0801, Japan}
\end{center}
\vs{20}

\centerline{{\bf{Abstract}}}
\vs{3}

The fuzzy algebra of $S^4$ is discussed by quantum deformation. To this end we embed the classical $S^4$ in the K\"ahler coset space $SO(5)/U(2)$. The harmonic functions of $S^4$ are constructed in terms of the complex coordinates of $SO(5)/U(2)$. Being endowed with the symplectic structure they can be deformed by the Fedosov formalism. We show that they generate the fuzzy algebra $\hat A_\infty (S^4)$ under the $\star$ product defined therein, by using the Darboux coordinate system. The fuzzy  spheres of higher even dimensions can be discussed similarly. 
 We give basic arguments for the generalization as well.

\vs{45}
\begin{flushleft}
\end{flushleft}

\end{titlepage}
\newpage

\renewcommand{\thefootnote}{\arabic{footnote}}
\setcounter{footnote}{0}

\section{Introduction}
\setcounter{equation}{0}

The fuzzy $4$-sphere was discussed in \cite{GKP} as the next simplest generalization of the $2$-sphere which had been extensively studied in the literature. In the string context the fuzzy spheres of four and other dimensions appeared as classical solutions\cite{M}\cite{MM} in the Matrix Model\cite{Matrix}. They represent non-flat $p$-branes in the string theory. It has been argued that perturbation around such classical solutions  provides us with non-commutative gauge  theories on the fuzzy spheres\cite{Ram2}\cite{pert}\cite{Kita}. The fuzzy spheres were discussed also as classical solutions of the DBI action which represent non-commutative backgrounds of $D$-string propagation\cite{BI}. Moreover  non-commutativity of spheres  was  found in the string theory with  the $AdS_n\times S^m$ geometry as well\cite{AdS}. 
 
\vspace{1cm}

All the above arguments  were developed for matrix realization of the fuzzy spheres. 
On the contrary in this paper we will discuss the fuzzy spheres by the Fedosov deformation quantization\cite{Fed}\cite{AM1}\cite{AM}\cite{AM3}, as long as their dimensions are even. The key point to this end was found 
in \cite{Ram2}\cite{Ram1}. Namely they  gave a proper account of  the relevance of  the coset space $SO(2a+1)/U(a)$   for the harmonics analysis of the fuzzy $S^{2a}$. 
Based on their findings  we will study quantum deformation of the harmonic functions of $S^{2a}$, exploiting the K\"ahler structure of the coset space $SO(2a+1)/U(a)$.

\vspace{1cm}

We shall briefly  review the works \cite{Ram2}\cite{Ram1}. 
The fuzzy $4$-sphere  is described by looking for an $N\times N$ matrix realization of the equation 
\begin{eqnarray}
\sum_{\mu=1}^5 x^\mu\cdot x^\mu = const.,  \label{al}
\end{eqnarray}
which classically describes the $4$-sphere. The matrices $x^\mu$ transform as ${\underline 5}$ of $SO(5)$. Hence   they are operators in an $N$-dimensional irreducible (spinor) representation of $SO(5)$. For a generic $N$ they generate an infinite dimensional  algebra under matrix multiplication. When ${\underline N}$ is the representation obtained by $n$-fold symmetric products of the spinor ${\underline 4}$ of $Spin(5)$, we have an arithmetical identity
\begin{eqnarray}
N^2 = {1\over 36}(n+1)^2(n+2)^2(n+3)^2 =\sum_{n \ge r_1 \ge r_2} D(r_1,r_2). \label{dim}
\end{eqnarray}
Here $D(r_1,r_2)$ is the dimension of the representation corresponding to the Young diagram of $SO(5)$, labelled by row length $(r_1,r_2)$.

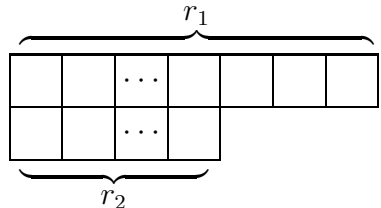
\setlength{\unitlength}{1mm}
\begin{picture}(70,28)(-40,-20)

\put(0,0){\line(1,0){49}}
\put(0,-7){\line(1,0){49}}
\put(0,-14){\line(1,0){28}}

\put(0,0){\line(0,-1){14}}
\put(7,0){\line(0,-1){14}}
\put(14,0){\line(0,-1){14}}
\put(21,0){\line(0,-1){14}}
\put(28,0){\line(0,-1){14}}

\put(35,0){\line(0,-1){7}}
\put(42,0){\line(0,-1){7}}
\put(49,0){\line(0,-1){7}}

\put(14,-7){\makebox(7,7){$\cdots$}}
\put(14,-14){\makebox(7,7){$\cdots$}}

\put(0,-2){
$\overbrace{\makebox(47,3){}  }^{\mbox{$r_1$}}$
}

\put(0,-15){
$\underbrace{\makebox(25,3){}  }_{\mbox{$r_2$}}$
}
\end{picture}

\noindent
For this special value of $N$,   products of the matrices $x^\mu$ generate a finite dimensional algebra which is isomorphic to the full set of $N\times N$ matrices. They are decomposed into sets of matrices which transform  irreducibly under $SO(5)$  according to the Young diagrams $(r_1,r_2)$ relevant in the sum. The matrix algebra is called $\hat A_n (S^4)$. From $\hat A_n (S^4)$ we may project out the generators corresponding to the Young diagrams with $r_2 \ne 0$ to define a subalgebra, called $ A_n (S^4)$. It is the classical analogue of the algebra generated by products of the harmonic functions on $S^4$, but clearly $ A_n (S^4)$ is no longer an associative algebra. The associativity is recovered at the limit $n\rightarrow \infty$.

\vspace{1cm}

In this paper we will reverse the above arguments. Namely we start with explicitly giving the harmonic functions $x^\mu$ of $S^4$. Then we will deform
  them according to the Fedosov formalism and realize the algebraic equation (\ref{al}) with the $\star$ product defined therein. However $S^4$ is a real $4$-dimensional manifold with no symplectic structure. As such the deformation quantization by Fedosov does not work for $S^4$. A hint to overcome this difficulty is to consider  a bundle over $S^4$ with fibre $S^2$, which is the  K\"ahler coset space 
$$
SO(5)/U(2) = \{SO(5)/SO(4)\}\{SO(4)/U(2)\} = S^4\times S^2.
$$
 Then $S^4$  may be described by the complex coordinate system of $SO(5)/U(2)$ , where a symplectic structure manifests. The  K\"ahler coset space $SO(5)/U(2)$  has a set of Killing potentials $M^A,\ A = 1,2,\cdots, 10$. By  the Lie-variation of the isometry $SO(5)$ they transform as $\underline {10}$:
\begin{eqnarray}
{\cal L_{R^A}} M^B =\sum_{C\in \underline {10}} f^{ABC}M^C, \label{1}
\end{eqnarray}
 and satisfy 
\begin{eqnarray}
 \sum_{A\in \underline {10}} M^AM^A = const.,  \label{2}
\end{eqnarray}
with $f^{ABC}$ the structure functions of $SO(5)$. The existence of such Killing potentials is known for the general K\"ahler coset space\cite{BW}. But an unusual feature of  $SO(5)/U(2)$  is  that from these Killing potentials one can construct a fundamental vector $x^\mu$ of $SO(5)$ by the tensor product
$
\underline {10}\otimes \underline {10} = \underline 5\oplus \cdots.
$  
By the same Lie-variation as above it transforms as 
\begin{eqnarray}
 {\cal L_{R^A}}\ x^\mu =\sum_{\nu\in \underline {5}} f^{A\mu\nu}\ x^\nu, \label{3}
\end{eqnarray}
in which  $f^{A\mu\nu}$  are matrix elements of the $SO(5)$-generators in the $5$-dimensional representation.
 We will then find  $x^\mu$ to obey the algebraic equation (\ref{al}).
The existence of such a fundamental vector 
 is characteristic for the class of the K\"ahler coset space $SO(2a+1)/U(a)$.
  In contrast with the matrix realization,  symmetric tensor products of $x^\mu$ 
  generate the commutative subalgebra of the harmonic functions of $S^4$, $A_\infty(S^4)$. 
  By the construction it is obvious that these harmonic functions are expressed by the complex coordinates of $SO(5)/U(2)$. Hence  
  they can now be  deformed by the Fedosov formalism to discuss the fuzzy $S^4$\cite{AM1}. 
  We will then examine 
  the fuzzy algebra  under  the Fedosov $\star$ product by taking  the Darboux coordinates\cite{AM}. It will be shown that 
 \begin{eqnarray}
 \sum_{\mu =1}^5 x^\mu\star x^\mu = d_0+ d_2 \hbar^2, \quad\quad\quad\quad
  [x^\mu,x^\nu]_\star = id_1\hbar \sum_{A\in \underline {10}} f^{A\mu\nu}M^A, \label{fuzzyal}
\end{eqnarray}
 with some constants $d_0, d_1$ and $d_2$. More generally we can show that the Fedosov $\star$ product of $x^\mu$  preserves the symmetry of $SO(5)$. Therefore repeating the $\star$ product generates the algebra isomorphic to $\hat A_\infty (S^4)$. 
 
\vspace{1cm}

The paper is organized as follows. In Section 2 we discuss the K\"ahler coset space $SO(5)/U(2)$. The Killing vectors, K\"ahler potential and the Killing potentials of the coset space are explicitly given. The fundamental vector $x^\mu$ of $SO(5)$ is constructed from the Killing potentials. In Section 3 we discuss  the harmonic functions of $S^4$. In Section 4   they are deformed by the Fedosov formalism in the  Darboux coordinates. They are shown to generate 
 the non-commutative  algebra $\hat A_\infty (S^4)$ under the $\star$ product defined therein. In Section 5 we explain the relation between the coset spaces $SO(5)/U(2)$ and $U(4)/U(3)\otimes U(1)$, which is useful to get better understanding of the former coset space. The whole arguments on the fuzzy $4$-sphere can be straightforwardly generalized to the case of the fuzzy $S^{2a}$. 
Appendix is devoted to give basic arguments for the generalization.

\section{The K\"ahler coset space $SO(5)/U(2)$}
\setcounter{equation}{0}

The coset space $SO(5)/U(2)$ is a K\"ahler manifold according to the Borel theorem\cite{Borel}.
 We shall study on an explicit construction of this manifold. The Lie-algebra of $SO(5)$ is given as
\begin{eqnarray}
 [t^{\mu\nu}, t^{\rho\sigma}] = i\delta^{\mu\rho}t^{\nu\sigma} -i\delta^{\nu\rho}t^{\mu\sigma}-i\delta^{\mu\sigma}t^{\nu\rho}+i\delta^{\nu\sigma}t^{\mu\rho}  \label{Lie}
\end{eqnarray}
where $t^{\mu\nu} = -t^{\nu\mu}$ with $\mu,\ \nu = 1,2,\cdots ,5$. We will decompose the generators $t^{\mu\nu}$ into the broken generators, denoted by $X^i$ and $X^{\bar i},\ i=1,2,3$ and the ones of the homogeneous group $U(2)$, denoted by $S^I, Y, \ I=1,2,3$:
\begin{eqnarray}
\{T^A\} \equiv \{X^{\bar i}, X^i,S^I,Y \}.  \label{decom}
\end{eqnarray}
 By noting the $SU(2)\otimes SU(2)$-subalgebra formed by
\begin{eqnarray}
S^1 &=& {1\over 2}(t^{23} + t^{14}),\ \ \ S^2 = {1\over 2}(t^{31} + t^{24}), \ \ \  S^3 = {1\over 2}(t^{12} + t^{34}),  \nonumber \\
P^1 &=& {1\over 2}(t^{23} - t^{14}),\ \ \ P^2 = {1\over 2}(t^{31} - t^{24}),  \ \ \  P^3 = {1\over 2}(t^{12} - t^{34}),  \nonumber 
\end{eqnarray}
they are identified as 
\begin{eqnarray}
X^1 &=& {1\over 2}(t^{15} + it^{25} ), \ \ \  
X^2 = {1\over 2}(-t^{35} + it^{45} ), \ \ \
X^3 = {1\over \sqrt 2 }( P^1 + iP^2), \nonumber \\
S^I &=& (S^1,S^2,S^3), \ \ \ Y = P^3. \nonumber 
\end{eqnarray}
In this basis the Casimir of $SO(5)$ takes  the form
\begin{eqnarray}
 T^A T^A = X^iX^{\bar i} + X^{\bar i}X^i+ S^-S^+ + S^+S^- + (S^3)^2 + (Y)^2,
 \nonumber
\end{eqnarray}
with $ S^{\pm} = {1\over \sqrt 2}(S^1 \pm iS^2)$. The non-trivial part of the Lie-algebra (\ref{Lie}) reads
\begin{eqnarray}
 [ Y, { X^1 \choose X^2 } ] &=& {1\over 2}{ X^1 \choose X^2 }, 
       \quad\quad\quad\quad  [Y, X^3] = X^3, \nonumber  \\  
\ [S^I, { X^1 \choose X^2 } ] &=& {1\over 2}\sigma^I { X^1 \choose X^2 }, 
       \quad\quad\quad [ S^I, X^3] = 0, \label{LLie}   \\
\ [X^1, X^{\bar 1} ] = {1\over 2}(Y + S^3), &\quad& 
  [X^2, X^{\bar 2} ] = {1\over 2}(Y - S^3), \quad
   [ X^3, X^{\bar 3} ] = Y  \nonumber \\
\ [X^1,X^{\bar 2} ] = {1\over \sqrt 2}S^+, &\quad& 
  [X^1,X^{\bar 3} ] = -{1\over \sqrt 2}X^{\bar 2}, \quad
  [X^2, X^{\bar 3}] = {1\over \sqrt 2}X^{\bar 1},  \nonumber \\
\ [X^1,X^2 ] = {1\over \sqrt 2}X^3, &\quad& 
  [X^1,X^3 ] = 0, \quad [X^2, X^3 ] = 0. \nonumber
\end{eqnarray}
The K\"ahler coset space $SO(5)/U(2)$ is parametrized by the coordinates corresponding to the broken generators $X^i$ and $X^{\bar i}$. 
From (\ref{LLie}) we find that under the homogeneous group $SU(2)$ the broken generators $X^1$ and $X^2$ transform as ${\underline 2}$, while $X^3$ as ${\underline 1}$. Therefore the K\"ahler coset space $SO(5)/U(2)$ is reducible. For an explicit construction we  have to be involved in the  general arguments given in \cite{Kugo}\cite{AS1}. But
 the homogeneous group contains a single $U(1)$ so that the construction is relatively easier.

\subsection{The Killing vectors}

First of all we discuss the holomorphic Killing vectors
 $ R^{A\alpha}(z)$ and $R^{A\ov \alpha}(\ov z)$ in the basis of the decomposition (\ref{decom}). 
 The standard application of the CCWZ formalism\cite{CCWZ} does not give the holomorphic Killing vectors $R^{A\alpha}$ satisfying the Lie-algebra (\ref{LLie}). Hence we extend the isometry group $SO(5)$ to the complex one $SO(5)^c$ and consider a coset space $SO(5)^c/\hat U(2)$ with the complex subgroup $\hat U(2)$ generated by $X^i, S^I, Y$\cite{Kugo}. As will be explicitly shown later, there is an isomorphism between this complex coset space $SO(5)^c/\hat U(2)$ and $SO(5)/U(2)$:
\begin{eqnarray} 
SO(5)/U(2) \cong SO(5)^c/\hat U(2).  \label{isore}
\end{eqnarray}
The holomorphic Killing vectors are obtained by applying the CCWZ formalism to the complex coset space  $SO(5)^c/\hat U(2)$. It is parametrized by complex coordinates $z^\alpha, \alpha = 1,2,3$ corresponding to the broken generators $X^i$. Consider a holomorphic quantity 
\begin{eqnarray}
\xi (z) = e^{z\cdot \bar X} \in SO(5)^c/\hat U(2)  \label{xi}
\end{eqnarray}
with 
$$
z\cdot \ov X = z^1 X^{\ov 1} + z^2 X^{\ov 2}+ z^3 X^{\ov 3}. \nonumber
$$
By left multiplication of $g = e^{i\epsilon^A T^A} \in SO(5)$ , we can find 
 the relation
\begin{eqnarray}
g\xi (z) = \xi (z') \hat h (z, g),  \label{trans}
\end{eqnarray}
appropriately choosing the holomorphic compensator 
 $\hat h (z, g) = e^{i\lambda (z,g)\cdot {\cal H}}  \in \hat U(2)$. Here 
 $\ep^A$ and $\lambda$s are global and local parameters parametrizing $g$ and $\hat h$ respectively as 
\begin{eqnarray}
\ep^A T^A &=& \ep^iX^{\bar i}+ \ep^{\bar i}X^i + \ep^-S^+ + \ep^+S^- + \ep^SS^3+ \ep^Y Y, \nonumber \\
\lambda \cdot {\cal H} &=& 
 \lambda^{\bar i}\cdot X^i +\lambda^- S^+ + \lambda^+ S^- + \lambda^S S^3 + 
 \lambda^Y Y.   \label{lambda}
\end{eqnarray}
This defines a holomorphic transformation of the coordinates $z^\alpha$ which realizes the isometry group non-linearly. When the  parameters $\epsilon^A$ are infinitesimal, (\ref{trans}) yields the holomorphic Killing vectors $R^{A\alpha}(z)$  as
\begin{eqnarray}
\delta z = z'^\alpha (z)  - z^\alpha = \epsilon^A R^{A\alpha}(z), \label{trans2}
\end{eqnarray}
which satisfy the Lie-algebra (\ref{LLie}).

\vspace{1cm}

To make the argument explicit we use the spinor representation of $SO(5)$. That is, the $SO(5)$-generators are given by 
\begin{eqnarray}
t^{\mu\nu} = -{i\over 2}\gamma^\mu \gamma^\nu,
\end{eqnarray}
with the $\gamma$-matrices 
\begin{eqnarray}
\gamma^i = i \left(
\begin{array}{cc}
   0 & \sigma^i  \\
   -\sigma^i & 0 
\end{array}
\right),  \quad\quad
\gamma^4 = \left(
\begin{array}{cc}
   0 & 1  \\
   1 & 0 
\end{array}
\right),  \quad\quad
\gamma^5 = \left(
\begin{array}{cc}
   -1 & 0  \\
   0 & 1
\end{array}
\right).   \nonumber
\end{eqnarray}
In the basis of the decomposition (\ref{decom}) the broken generators  become
\begin{eqnarray}
X^1 &=& -{i\over 4}(\gamma^1 + i\gamma^2)\gamma^5 = {1\over 2}\left(
\begin{array}{cccc}
   0 & 0 & 0 & 1  \\
   0 & 0 & 0 & 0  \\
   0 & 1 & 0 & 0  \\
   0 & 0 & 0 & 0  
\end{array}
\right),   \nonumber \\
X^2 &=& {i\over 4}(\gamma^3 - i\gamma^4)\gamma^5 = {1\over 2}\left(
\begin{array}{cccc}
   0 & 0 & 0 & 0  \\
   0 & 0 & 0 & 1  \\
   -1 & 0 & 0 & 0  \\
   0 & 0 & 0 & 0  
\end{array}
\right),  \label{gamma} \\
X^3 &=& -{1\over 4\sqrt 2}(\gamma^1 + i\gamma^2)(\gamma^3 - i\gamma^4) = {1\over \sqrt 2}\left(
\begin{array}{cccc}
   0 & 0 & 0 & 0  \\
   0 & 0 & 0 & 0  \\
   0 & 0 & 0 & 1  \\
   0 & 0 & 0 & 0  
\end{array}
\right),  \nonumber
\end{eqnarray}
while the ones of the homogeneous part 
\begin{eqnarray}
S^I = {1\over 2}\left(
\begin{array}{cc}
   \sigma^I & 0  \\
    0 & 0 
\end{array}
\right),  \quad\quad\quad
Y = {1\over 2}\left(
\begin{array}{cc}
   0 & 0  \\
    0 & \sigma^3
\end{array}
\right).   \nonumber
\end{eqnarray}
It is important to observe  algebraic relations of the formers such that
\begin{eqnarray}
 (X^i)^2 &=& 0, \quad\quad\quad i = 1,2,3,   \quad\quad\quad\quad\quad\quad
   \nonumber \\ 
  X^1X^2 &=& -X^2X^1 = {\sqrt 2 \over 4}X^3, \nonumber \\
  X^1X^3 &=& X^3X^1 = 0,\quad\quad X^2X^3 = X^3X^2 = 0. \nonumber 
\end{eqnarray}
Owing to these relations the holomorphic quantity (\ref{xi}) can be easily evaluated:
\begin{eqnarray}
\xi(z) = e^{z\cdot \ov X} = {1\over 2}\left(
\begin{array}{cccc}
   2 & 0 & -z^2 & 0  \\
   0 & 2 & z^1 & 0  \\
   0 & 0 & 2 & 0  \\
   z^1 & z^2 & z^3 & 2    
\end{array}
\right).  \label{expxi}
\end{eqnarray}
Choose the local parameters of the holomorphic compensator $\hat h (z, g)$ in (\ref{trans}) to be
\begin{eqnarray}
\lambda^{\bar 1} &=& \ov \ep^1 + {1\over \sqrt 2 }\ov \ep^3 z^2,\quad\quad
 \lambda^{\bar 2} = \ov \ep^2 - {1\over \sqrt 2 }\ov \ep^3 z^1,\quad\quad
 \lambda^{\bar 3} =  \ov \ep^3,  \nonumber \\
\lambda^+ &=& \sqrt 2  \ep^+ + {1\over 2\sqrt 2}\ov \ep^2 z^1 - {1\over 4}\ov\ep^3(z^1)^2,
   \quad\quad
\lambda^- =  \sqrt 2\ep^- + {1\over 2\sqrt 2}\ov \ep^1 z^2 -{1\over 4} \ov\ep^3(z^2)^2, \nonumber \\
\lambda^S &=&  \ep + {1\over 2}\ov \ep^1 z^1 - {1\over 2}\ov\ep^2 z^2 + {1\over 2\sqrt 2}\ov \ep^3 z^1z^2, \label{Y}  \\
\lambda^Y &=&  \ep^Y + {1\over 2}\ov \ep^1 z^1 + {1\over 2}\ov\ep^2 z^2 + \ov \ep^3 z^3. \nonumber
\end{eqnarray}
When the global parameters $\ep^A$ are infinitesimal, 
we find the holomorphic Killing vectors from the relations (\ref{trans}) and (\ref{trans2}) 
\begin{eqnarray}
&\quad& R^{\bar 1 1} = i, \quad\quad\quad\quad\quad  R^{11} = -{i\over 4}(z^1)^2, \nonumber 
 \\
&\quad& R^{\bar 2 1} = 0, \quad\quad\quad\quad\quad  R^{21} = {i\over 4}(2\sqrt 2 z^3-z^1z^2), \nonumber  \\
&\quad& R^{\bar 3 1} = 0, \quad\quad\quad\quad\quad  R^{31} = -{i\over 2}z^1z^3, \nonumber 
 \\
&\quad&R^{+1} = 0, \quad\quad R^{-1} = -{i\over \sqrt 2}z^2,  \quad\quad  R^{S1} =  -{i\over 2}z^1, \quad\quad R^{Y1} =  -{i\over 2}z^1,   \nonumber  \\
&\quad& R^{\bar 1 2} = 0, \quad\quad\quad\quad\quad  R^{12} = -{i\over 4}(2\sqrt 2 z^3+ z^1z^2), \nonumber 
 \\
&\quad&R^{\bar 2 2} = i, \quad\quad\quad\quad\quad  R^{22} = -{i\over 4}(z^2)^2, \label{kilvec}  \\
&\quad& R^{\bar 3 2} = 0, \quad\quad\quad\quad\quad  R^{32} = -{i\over 2}z^2z^3, \nonumber 
 \\
&\quad&R^{+2} = -{i\over \sqrt 2}z^1, \quad\quad R^{-2} = 0,  \quad\quad  R^{S2} =  {i\over 2}z^2, \quad\quad R^{Y2} =  -{i\over 2}z^2,   \nonumber \\
&\quad& R^{\bar 1 3} = -{i\sqrt 2\over 4} z^2,  \quad\quad\quad\quad\quad  R^{13} = -{i\over 4}z^1z^3, \nonumber 
 \\
&\quad& R^{\bar 2 3} = {i\sqrt 2 \over 4}z^1, \quad\quad\quad\quad\quad  R^{23} = -{i\over 4}z^2z^3, \nonumber  \\
&\quad& R^{\bar 3 3} = i, \quad\quad\quad\quad\quad  R^{33} = -{i\over 2}(z^3)^2, \nonumber 
 \\
&\quad& R^{+3} = 0, \quad\quad R^{-3} = 0,  \quad\quad  R^{S3} =  0, \quad\quad R^{Y3} = 
 -i z^3.   \nonumber 
\end{eqnarray}

\subsection{The K\"ahler potential}

Next we will discuss the K\"ahler potential of $SO(5)/U(2)$. We have recourse to the generalized CCWZ formalism adapted for the K\"ahler coset space\cite{Kugo}.   Consider a quantity
\begin{eqnarray}
 U(z,\ov z) \in SO(5)/U(2),
\end{eqnarray}
with $U^\dagger U = UU^\dagger = 1$.
But the standard parametrization of $U$, i.e, $U(z,\ov z) = e^{z\cdot \bar X - 
 \ov z \cdot X}$ does not give the metric of the type (1,1), i.e, $g_{\alpha\beta}= g_{\ov\alpha\ov\beta}= 0$. Therefore we employ the non-standard one, 
namely 
\begin{eqnarray}
 U(z,\ov z) = \xi (z)\zeta (z,\ov z),  \label{iso}
\end{eqnarray}
in which $\xi (z)$ is the holomorphic quantity defined by (\ref{xi}),
 while $\zeta (z,\ov z)$ an element of the complex subgroup $\hat U(2)$. We parametrize the latter as 
\begin{eqnarray}
 \zeta (z,\ov z)= e^{a(z,\ov z)\cdot X} e^{b(z,\ov z)\cdot S} e^{c(z,\ov z)Y},
  \label{para}
\end{eqnarray}
with $a\cdot X = a^{\bar i}X^i$ and $b\cdot S =  b^IS^I$. 
Here $a^{\bar i}$ are complex functions, while $b(z,\ov z)$ and $c(z,\ov z)$ are chosen to be real functions because the purely imaginary parts can be absorbed into an element of $H$. They are determined by the unitary condition $U^\dagger U = 1$ which reads 
\begin{eqnarray}
 \xi^\dagger (\ov z) \xi (z) = e^{-\bar a (z,\bar z)\cdot \bar X}
                                      e^{-2b (z,\bar z)\cdot  S}
                                      e^{-2 c (z,\bar z)\cdot Y}
                                      e^{-a (z,\bar z)\cdot X}.
   \label{unit}
\end{eqnarray}
We then remark that (\ref{iso}) is an concrete expression of the isomorphism (\ref{isore}) between the coset spaces $SO(5)/U(2)$ and $SO(5)^c/\hat U(2)$. In ref. \cite{Kugo} it was shown that we may identify the local parameter $c(z,\ov z)$ to be the K\"ahler potential of the manifold\begin{eqnarray}
-2c(z,\ov z) = K(z,\ov z),
\end{eqnarray}
because  the transformation  (\ref{trans2}) induces the change
\begin{eqnarray}
c(z,\ov z) \rightarrow c(z,\ov z) + {i\over 2}(\lambda^Y (z) - \ov \lambda^Y (\ov z)),
 \label{ktrans}
\end{eqnarray}
in (\ref{unit}). Here $\lambda^Y (z)$ and $\ov \lambda^Y (\ov z)$ are the holomorphic functions given in (\ref{Y}).

\vspace{1cm}

We will apply this argument to find an explicit form of the K\"ahler potential for  $SO(5)/U(2)$. It is again convenient to work out in  the spinor representation (\ref{gamma}). 
The $(3,3)$-element of the $r.h.s.$ in the unitary condition (\ref{unit}) reduces to 
\begin{eqnarray}
 [\xi^\dagger (\ov z)\xi ]_{33} = [e^{-2c(z,\ov z)Y}]_{33} = e^{-c(z,\ov z)}.  \nonumber 
\end{eqnarray}
By calculating the  $l.h.s.$  with (\ref{expxi}) it  yields 
\begin{eqnarray}
K(z,\ov z) = 2\log  (1 + {1\over 4}|z^1|^2 + {1\over 4}|z^2|^2 + 
{1\over 2}|z^3|^2).   \label{kae}
\end{eqnarray}
We may check  the transformation property of this K\"ahler potential by the Killing vectors (\ref{kilvec}). It   indeed changes  as (\ref{ktrans}) with the holomorphic function $\lambda^Y (z)$ given in  (\ref{Y}). We observe that the form of the K\"ahler potential  is almost the same as the one of $CP^4(=U(4)/U(3)\otimes U(1))$. But  the isometries realized on both manifolds are clearly different. We will later come back to inspect a relationship between them.

\subsection{Killing potentials}

Finally we calculate the Killing potentials $M^A(z,\ov z)$ for $SO(5)/U(2)$. According to ref. \cite{BW} they are given by 
\begin{eqnarray}
-iM^A = K_{,\alpha} R^{A\alpha} - F^A.  \label{form}
\end{eqnarray}
Here $F^A$ follow from the transformation property (\ref{ktrans}) of the K\"ahler potential, i.e., 
$$
-i\lambda^Y = \epsilon^AF^A. 
$$
By using $\lambda^Y$ given in (\ref{Y}) together with (\ref{kilvec}) and (\ref{kae}), we calculate the $r.h.s.$ of (\ref{form}) to obtain the the Killing potentials $M^A(z,\ov z)$
\begin{eqnarray}
M^1 &=& -{1\over 2f}(z^1 -{1\over \sqrt 2}\ov z^2 z^3 ) , \quad\quad\quad\quad\quad c.c.,
  \nonumber \\
M^2 &=& -{1\over 2f}(z^2 +{1\over \sqrt 2}\ov z^1 z^3 ) , \quad\quad\quad\quad\quad c.c.,
  \nonumber \\
M^3 &=& -{1\over f}z^3,  \quad\quad\quad\quad\quad c.c.,
  \label{k1} \\
M^+ &=& {1\over 2\sqrt 2 f}  \ov z^2 z^1, \quad\quad\quad\quad M^- = {1\over 2 \sqrt2 f }\ov z^1 z^2,
  \nonumber \\
M^S &=& {1\over 4f}(|z^1|^2 - |z^2|^2 ),   \quad\quad\quad\quad
M^Y = -{1\over 2f}(2 - |z^3|^2),  \nonumber
\end{eqnarray}
in which
$$
 f = 1 + {1\over 4}|z^1|^2 + {1\over 4}|z^2|^2 +{1\over 2}|z^3|^2. 
$$
From these Killing potentials we calculate the fundamental vector $x^\mu$ by the formula\begin{eqnarray}
x^\mu= {1\over 8} \varepsilon^{\mu\nu\rho\sigma\delta}M^{\nu\rho}M^{\sigma\delta},  \label{ff} 
\end{eqnarray}
with $\varepsilon^{\mu\nu\rho\sigma\delta}$ the totally antisymmetric tensor of $SO(5)$. 
It reads 
\begin{eqnarray}
 {1\over 2}(-ix^1 + x^2) &=& 
 \sqrt 2M^3M^{\bar 2} - \sqrt 2M^+M^{2} - M^{ 1} (M^S - M^Y),
 \hspace{0.5cm} c.c.,  \nonumber \\
 {1\over 2}(ix^3 + x^4) &=& 
 -\sqrt 2M^3M^{\bar 1} - \sqrt 2M^-M^1 + M^2 (M^S + M^Y),
\hspace{0.5cm} c.c.,  \label{444} \\
x^5 &=& (M^S)^2 - (M^Y)^2 + 2M^+M^- -2 M^3M^{\bar 3}. \nonumber
\end{eqnarray}
We then find 
\begin{eqnarray}
-ix^1 + x^2 &=& {1\over f}(z^1 +{1\over \sqrt 2}\ov z^2 z^3 ), \quad\quad\quad\quad\quad c.c.,   \nonumber \\
ix^3 + x^4 &=& {1\over f}(z^2 -{1\over \sqrt 2}\ov z^1 z^3 ),  \quad\quad\quad\quad\quad c.c., \label{quintet} \\
x^5 &=& {1\over f}({|z^1|^2\over 4} + {|z^2|^2\over 4} -{|z^3|^2 
\over 2}-1). \nonumber
\end{eqnarray}
The respective transformation properties (\ref{1}) and (\ref{3}) of $M^A$ and $x^\mu$ are obvious by the construction. On the other hand the algebraic equation (\ref{2}) and (\ref{al}) follow from the theorem given in ref. \cite{AM3}. But we have here checked them by direct calculations:
\begin{eqnarray}
M^AM^A = 1,   \quad\quad\quad  x^\mu x^\mu = 1.  \label{cc}
\end{eqnarray}

\section{The harmonic functions of $S^4$}
\setcounter{equation}{0}

We now show that the fundamental vector $x^\mu$ generates harmonic functions of $S^4$. Define  a $`` false "$ metric of $SO(5)/U(2)$ by 
\begin{eqnarray}
{\mathop{g}^\circ}\ ^{\alpha \bar\beta} &\equiv&
R^{A\alpha} R^{A\bar\beta}, \nonumber \\
{\mathop{g}^\circ}\ ^{\alpha\beta} &\equiv& 
R^{A\alpha}R^{A\beta}, \quad\quad\quad
{\mathop{g}^\circ}\ ^{\bar\alpha \bar\beta} \equiv
R^{A\bar\alpha}R^{A\bar\beta}. \label{metric}
\end{eqnarray}
They satisfy the Killing equations 
\begin{eqnarray}
{\cal L}_{R^A}{\mathop{g}^\circ}\ ^{\alpha \bar\beta} =0, \quad\quad etc.. \nonumber
\end{eqnarray}
By (\ref{kilvec}) we find that 
\begin{eqnarray}
{\mathop{g}^\circ}\ ^{\alpha\beta} &=& 0, \quad\quad\quad c.c., \label{metricc}
\end{eqnarray}
and 
$\displaystyle{\mathop{g}^\circ}\ ^{\alpha \bar\beta}$ is given by
\begin{eqnarray}
{\mathop{g}^\circ}\ ^{1 \bar 1} &=& 
   1 + {1\over 2}|z^1|^2 + {1\over 2}|z^2|^2 +{1\over 4}|z^1|^2|z^3|^2 \nonumber \\
   & &   + {1\over 16}|z^1|^4 + {1\over 16}(2\sqrt 2 z^3 -z^1z^2)(2\sqrt 2 \bar z^3
    -\bar z^1\bar z^2 ),   \nonumber    \\
 {\mathop{g}^\circ}\ ^{2 \bar 2} &=& 
   1 + {1\over 2}|z^1|^2 + {1\over 2}|z^2|^2 +{1\over 4}|z^2|^2|z^3|^2 \nonumber \\
   & &   + {1\over 16}|z^2|^4 + {1\over 16}(2\sqrt 2 z^3 +z^1z^2)(2\sqrt 2 \bar z^3
    +\bar z^1\bar z^2 ),   \nonumber   \\
{\mathop{g}^\circ}\ ^{3 \bar 3} &=& 
 (1 + {1\over 2}|z^3|^2 )(1 + {1\over 8}|z^1|^2 + {1\over 8}|z^2|^2 + 
   {1\over 2}|z^3|^2 ), \nonumber \\
{\mathop{g}^\circ}\ ^{1 \bar 2} &=& 
  {\sqrt 2 \over 8}((z^1)^2\bar z^3 - (\bar z^2)^2 z^3) 
  + z^1\bar z^2 ({1\over 16}|z^1|^2 + {1\over 16}|z^2|^2 + {1\over 4}|z^3|^2 ), \nonumber \\
{\mathop{g}^\circ}\ ^{1 \bar 3} &=& 
 -{\sqrt 2\over 4}\bar z^2(1 + {1\over 2}|z^3|^2) + 
z^1\bar z^3 ({1\over 16}|z^1|^2 + {1\over 16}|z^2|^2 +{1\over 4}|z^3|^2 + {1\over 2} ), \nonumber \\
{\mathop{g}^\circ}\ ^{2 \bar 3} &=&
 {\sqrt 2\over 4}\bar z^1(1 + {1\over 2}|z^3|^2) + 
z^2\bar z^3 ({1\over 16}|z^1|^2 + {1\over 16}|z^2|^2 + {1\over 4}|z^3|^2 + {1\over 2}), \nonumber
\end{eqnarray}
and their complex conjugates. 
Therefore 
 they give a $(1,1)$ metric, but 
the $\displaystyle{\mathop{g}^\circ}\ ^{\alpha \bar\beta}$ is not the inverse of $g_{\alpha\bar\beta}$ obtained from the K\"ahler potential (\ref{kae}).
 This discrepancy comes from  the fact that
$$
{\mathop{g}^\circ}\ ^{\alpha \bar\beta}\mathop{|}_{z=\ov z =0} \ne g^{\alpha \bar\beta}\mathop{|}_{z=\ov z =0} .
$$
It is a quite general phenomenon when the K\"ahler coset space is reducible. The correct inverse metric is given by 
\begin{eqnarray}
g^{\alpha \bar\beta} = R^{A\alpha}(UPU^{-1})^{AB}R^{B\bar \beta}, \label{U}
\end{eqnarray}
as well as
$$
g^{\alpha \beta} = R^{A\alpha}(UPU^{-1})^{AB}R^{B \beta}= 0,\quad\quad\quad
 c.c..
$$
Here $U $ is the quantity defined by (\ref{iso}), but in the adjoint representation. 
$P$ is a matrix which  has  non-vanishing elements only in the  diagonal blocks corresponding to the broken generators $X^a = (X^{\ov i},X^i )$ such that 
\begin{eqnarray}
 P^{i\ov j} = P^{\ov j i} =
\left(
\begin{array}{ccccc}
2 & 0 &  0  \\ 
 0   & 2 &  0  \\  
  0   & 0 & 1    \\
\end{array}\right). \nonumber
\end{eqnarray}
(For the details on this point the readers can refer to \cite{AS1}\cite{AM4}.) 
Nonetheless the Laplacian on $SO(5)/U(2)$ with the $`` false "$ metric is a nice property , namely the Laplacian for scalar fields is given by 
\begin{eqnarray}
\Delta &=& {1\over \displaystyle\sqrt {\mathop{g}^\circ}}
\partial_\alpha ( \sqrt {\mathop{g}^\circ}\ {\mathop{g}^\circ}\ ^{\alpha\bar\beta}\partial_{ \bar \beta}) + c.c.
 \nonumber \\
&=&  (R^{A\alpha}\partial_\alpha + R^{A\bar\alpha}\partial_{\bar\alpha}) (R^{A\beta}\partial_\beta + R^{A\bar\beta}\partial_{\bar\beta}) = {\cal L}_{R^A}{\cal L}_{R^A}.
 \label{lap}
\end{eqnarray}
Here $\displaystyle{\mathop{g}^\circ}=(\det \displaystyle{\mathop{g}^\circ}_{\alpha \bar\beta})^2$.  
It can be easily shown by using (\ref{metric}) with (\ref{metricc}) and the formulae following from them:
$$
R^{A\alpha}_{\ \ \ ,\alpha}R^{A\beta} = 0, \quad\quad\quad\quad
 R^{A\alpha}_{\ \ \ ,\alpha}R^{A\bar \beta} = -{1\over \sqrt {\mathop{g}^\circ}}
 \partial_\alpha (\sqrt {\mathop{g}^\circ})\ {\mathop{g}^\circ}\ ^{\alpha\bar\beta}.
$$
Act the Laplacian on  the fundamental vector $x^\mu$ given by (\ref{quintet}).
 Owing to (\ref{3}) we find
\begin{eqnarray}
\Delta x^\mu =  {\cal L}_{R^A}{\cal L}_{R^A} x^\mu = f^{A\mu\nu}f^{A\nu\rho} x^\rho  = c_{\underline 5} x^\mu,    \label{lapp}
\end{eqnarray}
in which  $c_{\underline 5}$ is   the  Casimir of $SO(5)$ in ${\underline 5}$ taking the value  $-2$ 
in the basis  given by (\ref{LLie}). This equation  implies that the fundamental vector $x^\mu$ is an 
eigenvector of the Laplacian (\ref{lap}) and  gives a basis of the harmonic functions of $S^4$. 

\vspace{1cm}

Note that 
 the Killing potentials $M^A$ given by (\ref{k1}) are also eigenvectors of the Laplacian (\ref{lap}), {\it i.e.,} 
\begin{eqnarray}
\Delta M^B = {\cal L}_{R^A}{\cal L}_{R^A} M^B = f^{ABC}f^{ACD}M^D = c_{\underline {10}} M^B,  \label{lappp}
\end{eqnarray}
with $c_{\underline {10}} = -3$  which is  the Casimir of $SO(5)$ in ${\underline {10}}$. But $M^A \notin  A_\infty(S^4)$. In other words the Killing potentials $M^A$
 are not harmonic functions of $S^4$, but of $SO(5)/U(2)$, since they cannot be obtained by symmetric products of $x^\mu$.

\section{Fuzzy algebrae}
\setcounter{equation}{0}

The Fedosov formalism\cite{Fed} for the deformation quantization  provides us with 
 $\star$ product of functions on symplectic manifolds. It may be applied for the K\"ahler manifold most effectively as shown in \cite{AM1}. When we change the coordinates $(z^\alpha,\bar z^{\alpha})$ to  $(q^\alpha,p_\alpha )$ as
\begin{eqnarray}
q^\alpha = z^\alpha, \quad\quad\quad p_\alpha = -iK_{,\alpha}, \label{Dar}
\end{eqnarray}
with the K\"ahler potential, 
 the K\"ahler two-form can be put in the form
\begin{eqnarray}
d\omega = dp_\alpha \wedge dq^\alpha.   \nonumber
\end{eqnarray}
Hence $(q^\alpha,p_\alpha )$ are the Darboux coordinates. 
Then the Fedosov $\star$ product for the K\"ahler manifold 
  reduces to 
\begin{eqnarray}
 a(p,q)\star b(p,q) =
                 \sum_n {1\over n!} a(p,q)[ {i\hbar \over 2}
                             ({\overleftarrow \partial \over \partial p_\alpha }
                             {\overrightarrow \partial \over \partial  q^\alpha}
                           -  {\overrightarrow\partial \over \partial p_\alpha }
                             {\overleftarrow \partial \over \partial q^\alpha})]^n b(p,q),
                         \label{Moy}
\end{eqnarray}
which is the Moyal product\cite{AM}. The Killing potentials are given by  (\ref{form}) for the general K\"ahler coset space. In terms of the Darboux coordinates 
 the Killing potentials become 
\begin{eqnarray}
-iM^A(q,p) = ip_\alpha R^{A\alpha}(q) - F^A(q).  \label{formm} 
\end{eqnarray}
In ref. \cite{AM3} it was shown that with the $\star$ product (\ref{Moy}) they satisfy the fuzzy algebrae
\begin{eqnarray}
 [ M^A, M^B ]_\star
 &=& -i \hbar f^{ABC} M^C,  \nonumber  \\
  M^A\star M^A &=&  c_0 + c_2\hbar^2,  \nonumber
\end{eqnarray}
in which $c_0$ and $c_2$ are constants. For $SO(5)/U(2)$ we find that 
$$
c_0 = 1 \quad\quad\quad\quad c_2 = -1
$$
by the normalization of the $SO(5)$-algebra in (\ref{LLie}). 
The fundamental vector $x^\mu$ is also expressed by the 
 Darboux coordinates, owing to the formula (\ref{444}). Plugging the Killing potentials (\ref{formm}) into the formula we find $x^\mu$ to take the simple form
\begin{eqnarray}
ix^1 + x^2 &=& {1\over 2}(4ip_1 + \sqrt 2 ip_3 q^2),    \nonumber \\
-ix^1 + x^2 &=& -{1\over 2}[\ ip_1(q^1)^2 + ip_2q^1q^2 + ip_3q^1q^3 -2\sqrt 2 ip_2q^3 -2 q^1 \ ], \nonumber \\
ix^3 + x^4 &=& -{1\over 2}[\ ip_1q^1q^2 + ip_2(q^2)^2 + ip_3q^2q^3 +2\sqrt 2 ip_1q^3 - 2q^2 \ ],  \label{qq} \\
-ix^3 + x^4 &=& {1\over 2}(4ip_2-\sqrt 2 ip_3q^1 ), \nonumber \\
x^5 &=& ip_1q^1 + ip_2q^2 - 1. \nonumber
\end{eqnarray}
Then the fuzzy algebrae (\ref{fuzzyal}) can can be easily checked. We find that 
\begin{eqnarray}
 x^\mu\star x^\mu = 1 - \hbar^2, \quad\quad\quad\quad
 \   [x^\mu,x^\nu]_\star =  -2i\hbar f^{A\mu\nu}M^A.  \label{scalar}
\end{eqnarray}
The coefficients $f^{A\mu\nu}$ should be  matrix elements of the $SO(5)$-generators in the 5-dim\-ensional representation 
because the Jacobi identity of the commutator with the $\star$ product.
 The $\star$ product (\ref{Moy}) preserves the symmetry of $SO(5)$. Namely we have 
$$
{\cal L}_{R^A}\left({\overleftarrow \partial \over \partial q^\alpha}{\overrightarrow\partial \over \partial p_\alpha}\right) = {\cal L}_{R^A}\left({\overleftarrow \partial \over \partial z^\alpha} ig^{\alpha\bar\beta}{\overrightarrow\partial \over z^{\bar\beta} }\right) = 0,
$$
due to the Killing equation
$
{\cal L}_{R^A}g^{\alpha\bar\beta} = 0,
$
in which $g^{\alpha\bar\beta}$ is the inverse metric of $g_{\alpha\bar\beta}$ or equivalently given  by (\ref{U}). Therefore the symmetric product $\{x^\mu,x^\nu\}_\star$ transforms as a tensor of the second rank. Subtracting the scalar component from this product by (\ref{scalar}) one obtains the harmonic function in $\underline {14}$ of $SO(5)$. Thus repeating the symmetric or antisymmetric $\star$ product generates 
the fuzzy algebra $\hat A_\infty(S^4)$.

\section{Relation between $SO(5)/U(2)$ and $U(4)/U(3)\otimes U(1)$}
\setcounter{equation}{0}

As has been noted at the beginning of Section 2 the K\"ahler coset space $SO(5)/U(2)$ is reducible, but $U(4)/U(3)\otimes U(1)(\cong SO(6)/U(3))$ not. 
We will discuss on a relation between these K\"ahler coset spaces. The 
Lie-algebra of $U(4)$ are given by 
\begin{eqnarray}
 [T_I^J, T_K^L ] = -\delta_K^J T_I^L + \delta_I^L T_K^J, \nonumber
\end{eqnarray}
where $(T_I^J)^\dagger = T^I_J, I,J = 1,2,3,4.$ Under the subgroup $U(3)$ the generators $T_I^J$ are decomposed as
\begin{eqnarray}
\{T_I^J\} =\{T^4_i, T^i_4, T_i^j, T^4_4 \}, \quad\quad i,j = 1,2,3. \nonumber
\end{eqnarray}
 We parametrize the K\"ahler coset space $U(4)/U(3)\otimes U(1)$  by the coordinates  $\phi^\alpha$ and $\bar \phi^{\alpha}, \ \alpha = 1,2,3$, which respectively correspond to the broken generators $ T^4_i$ and $T^i_4$. 
 Then the Killing vectors $R^{J\ \alpha}_I(\phi )$ and the complex conjugates are given by 
\begin{eqnarray}
R^{4\ \alpha}_i &=& i\delta_i^{\alpha},  \quad\quad\quad\quad\quad R^{i \ \alpha}_4 = -i\phi^i \phi^\alpha,  \nonumber \\
 R^{i\ \alpha}_j &=& -i\delta^\alpha_j\phi^i, \quad\quad\quad\quad\quad 
 R^{4\ \alpha}_4 = i\phi^\alpha.    \label{cpkill}
\end{eqnarray}
 The K\"ahler potential $\tilde K(\phi,\ov \phi )$ and the Killing potentials $M_I^J(\phi,\ov \phi )$ of $U(4)/U(3)\otimes U(1)$ respectively are found to take the forms
\begin{eqnarray}
\tilde K = \log (1 +|\phi^1|^2 + |\phi^2|^2 +|\phi^3|^2 ) \equiv \log \tilde f, \nonumber
\end{eqnarray}
and
\begin{eqnarray}
M^1_4 &=& -{1 \over \tilde f}\phi^1,  \quad\quad\quad\quad\quad c.c.,   \nonumber \\
M^2_4 &=& -{1 \over \tilde f }\phi^2,  \quad\quad\quad\quad\quad c.c., \label{kilU} \\
M^3_4 &=& -{1\over \tilde f }\phi^3 ,  \quad\quad\quad\quad\quad c.c.,   \nonumber \\
M^i_j &=& {1 \over \tilde f }\phi^i\bar \phi^{ j}, \quad\quad\quad\quad\quad
M^4_4 = {1\over \tilde f}.   \nonumber 
\end{eqnarray}
From (\ref{cpkill}) we find that
\begin{eqnarray}
 {\mathop{\tilde g}^\circ}\ ^{\alpha \bar\beta} \equiv R^{J\ \alpha}_I \overline{R^{J\ \beta}_I } &=& \tilde f (\delta^{\alpha\bar \beta} +  \phi^\alpha \bar\phi^{\beta}),   \nonumber \\
 {\mathop{\tilde g}^\circ}\ ^{\alpha \beta} \equiv R^{J\ \alpha}_I R^{I\ \beta}_J &=& 0. \quad\quad\quad\quad c.c..\label{metric2}
\end{eqnarray}
On the contrary to the case  of $SO(5)/U(2)$ it gives the correct inverse metric of $\tilde K_{,\alpha\bar \beta}$. This is a fact which always holds when the K\"ahler coset space is irreducible.

\vspace{1cm}

The isometry group $U(4)$ contains $SO(5)$. Hence the generators $T_I^J$ are decomposed  also under this subgroup as
$
{\underline 16} \rightarrow {\underline {10}} + {\underline 5} + {\underline 1}. \nonumber
$
They  are grouped into 
\begin{eqnarray}
&\underline {10}&: \left\{
\begin{array}{ll}
X^1 = {1\over 2}(T^1_4 + T^3_2),  \quad\quad\quad\quad h.c.,&  \\
   &  \\
X^2 = {1\over 2}(T^2_4 - T^3_1),  \quad\quad\quad\quad h.c., &  \\
&  \\
X^3 =  {1\over \sqrt 2 }T^3_4,  \quad\quad\quad\quad  h.c., &  \\
&  \\
S^+ = {1\over \sqrt 2 } T^1_2, \quad\quad S^- = {1\over \sqrt 2 }T^2_1, \quad\quad 
S^3 = {1\over 2}(T^1_1 - T^2_2),  &    \\
&  \\
Y =  {1\over 2}(T^3_3 - T^4_4), &  
\end{array}\right.    \nonumber \\
      & \quad &     \nonumber    \\
&{\underline 5}&: \left\{
\begin{array}{ll}
{1\over 2}(T^1_4 - T^3_2), &  \quad\quad\quad\quad h.c., \\
&  \\
 {1\over 2}(T^2_4 + T^3_1), &   \quad\quad\quad\quad h.c.,\\
 &  \\
  {1\over 2\sqrt 2}(-T^1_1 - T^2_2 +T^3_3 + T^4_4),  &  
\end{array}\right.   \label{norm}  \\
     & \quad &     \nonumber    \\
   &\underline {1}&:  \quad {1\over 2\sqrt 2}(T^1_1 + T^2_2 +T^3_3 + T^4_4).  \nonumber 
\end{eqnarray}
The generators in $\underline{10}$  satisfy the $SO(5)$-Lie-algebra in the form (\ref{LLie}). 
Correspondingly the Killing potentials (\ref{kilU}) are decomposed to yield those of $SO(5)/U(2)$ and the fundamental vector, respectively given by (\ref{k1}) and (\ref{quintet}).
  For the  precise identification we should understand the scaling
$$
\phi^1 = {1\over 2}z^1, \quad\quad\quad \phi^2 = {1\over 2}z^2, \quad\quad\quad 
\phi^3 = {1\over \sqrt 2}z^3, \quad\quad\quad \tilde K = {1\over 2}K.
$$
(Note also a slight difference between the normalization of the fundamental vector (\ref{quintet}) and that  of the corresponding generators   (\ref{norm}).)
 The Killing vectors of  $SO(5)/U(2)$, given by (\ref{kilvec}), can be obtained by similarly decomposing those given by (\ref{cpkill}).

\vspace{1cm}

The Killing potentials of $U(4)/U(3)\otimes U(1)$ may be decomposed under any other subgroup. The unusual feature of the decomposition under $SO(5)$ is that the Killing potentials in $\underline {10}$ and $\underline {5}$ each obey the constraints
(\ref{cc}). 
For a representation ${\underline n}$ of a generic subgroup 
for the isometry group $U(4)$ we find that
$$
 \mathop\sum_{{J \choose I} \in {\underline n}} M^J_IM^I_J \ne const..
$$
For instance,  take  a set of $M_i^j, i,j =1, 4$.  The corresponding set of the Killing vectors $R_i^{j\ \alpha}$ is a non-linear realization of the subgroup $U(2)$ generated by $T_i^j, i,j =1,4$. By the Lie-variation with respect to them $M_i^j$ transform as the adjoint representation of $U(2)$. However we find that 
$$
\sum_{i,j = 1,4} M_i^jM^i_j = {1\over \tilde f^2 }(1+|\phi^1|^2)^2   \ne const.,
$$
and 
\begin{eqnarray}
\sum_{i,j = 1,4}R_i^{j\ \alpha} \overline {R_i^{j\ \beta}} &=& (1+|\phi^1|^2)(\delta^\alpha_1\delta^{\bar \beta}_{\bar 1} + \phi^\alpha \bar\phi^{ \beta}), \nonumber \\
\sum_{i,j = 1,4}R_i^{j\ \alpha}R_j^{i\ \beta} &=& -( \phi^\alpha -\delta^\alpha_1\phi^1)(\phi^\beta- \delta^\beta_1 \phi^1 ).  \nonumber
\end{eqnarray}
We might say that the last two equations define a {\it false} metric of some manifold. But  it is degenerate at $\phi^\alpha = \bar\phi^{\alpha} = 0$, and is no longer of $(1,1)$ type.

\section{Conclusions}
\setcounter{equation}{0}

One of the important ingredients of this paper is that  we have found  the harmonic functions of $S^4$  in the form (\ref{quintet}). For this purpose we considered a bundle over $S^4$ with fibre $S^2$, which is the K\"ahler coset space $SO(5)/U(2)$. We have constructed the Killing potentials for $SO(5)/U(2)$ as (\ref{k1}). The harmonic functions  (\ref{quintet}) followed from them by  the formula (\ref{ff}). Hence they were expressed by the complex coordinates $z^\alpha$ and 
$\bar z^\alpha, \alpha= 1,2,3$ of the K\"ahler coset space $SO(5)/U(2)$. 

\vspace{1cm}

We can apply the deformation quantization by Fedosov\cite{Fed}\cite{AM1} for those harmonic functions and explore the fuzzy $S^4$ with the $\star$ product defined therein. 
To do this most conveniently we changed the complex coordinates  $(z^\alpha,\bar z^{\alpha})$ to the Darboux coordinates $(q^\alpha, p_\alpha)$ defined by (\ref{Dar}). Then  the Fedosov $\star$ product reduced to the usual Moyal product (\ref{Moy}), and the deformation quantization was much simplified. Moreover the harmonic functions of $S^4$  (\ref{quintet}) were readily expressed in the Darboux coordinates as (\ref{qq}). It consists of another important  ingredient of this paper. As the result  we were able to easily show the fuzzy algebrae (\ref{scalar}).

\vspace{1cm}

 In \cite{GKP}\cite{MM}  it was discussed that $N\times N$ matrix obeying the constraint (\ref{al}) generates the matrix algebra $\hat A_n(S^4)$, when $N$ takes the special value such  as (\ref{dim}) given by an integer $n$. Symmetric traceless products of the matrices up to order $ n$  form its subgroup $ A_n(S^4)$, which is not associative. In the limit $n \rightarrow \infty$ the associativity is recovered and $A_\infty(S^4)$ becomes the algebra equivalent to the one generated by the commutative products of the harmonic functions. We have shown that this commutative algebra of the harmonic functions becomes the non-commutative one $\hat A_\infty(S^4)$ by the deformation quantization by Fedosov. This is the main result of this paper.

\vspace{1cm}

These arguments on $S^4$ can be straightforwardly generalized to the case of $S^{2a}$. This time we consider a bundle over $S^{2a}$ with fibre $SO(2a)/U(a)$, which is the K\"ahler coset space $SO(2a+1)/U(a)$.  In Appendix we show an explicit way to construct the harmonic functions  of $S^{2n}$ in 
 the symplectic coordinates of the K\"ahler coset space $SO(2a+1)/U(a)$. Although we do not discuss in details, it is obvious that we can find the fuzzy algebra  $\hat A_\infty(S^{2n})$ by applying the deformation quantization for those harmonic functions similarly to the case of $S^4$.

\vspace{1cm}

 The  arguments in this paper were done by  fully  exploiting the the K\"ahler structure of $SO(5)/U(2)$. It was noticed in \cite{Kita}\cite{Nair} that the K\"ahler structure  is important for studying the Matrix Model on some non-commutative coset spaces. It is desired to extend their study to non-commutative backgrounds with the general K\"ahler coset space geometry  following the works\cite{AM3}.

\vspace{2cm}
\noindent
{\Large\bf Acknowledgements}

T.M. would like to thank 
Y. Kitazawa and Y. Kimura for discussions. 
The work of S.A. was supported in part  by the Grant-in-Aid for Scientific Research No.
13135212.

\vspace{3cm}

\appendix

\apsect{$S^{2a}$ in $SO(2a+1)/U(a)$  }
\setcounter{equation}{0}

The $2a$-sphere  is described by the coordinates of the K\"ahler coset space $SO(2a+1)/U(a)$ as noted by
\begin{eqnarray}
SO(2a+1)/U(a) &=&\{SO(2a+1)/SO(2a)\}\times \{SO(2a)/U(a)\}  \nonumber \\
 &=& S^{2a}\times SO(2a)/U(a). \nonumber
\end{eqnarray}
 To show this it suffices to explicitly construct  $SO(2a+1)/U(a)$. 
 It is a reducible K\"ahler coset space. The direct construction following the arguments in Section 2 is rather involved. Instead we will do it via the irreducible K\"ahler coset space $SO(2a+2)/U(a+1)$, as was done in Section 5. 
 $SO(2a+2)/U(a+1)$ may be constructed according to the general method for the irreducible K\"ahler coset space discussed in refs \cite{AM}. The generators of $SO(2a+2)$
 are decomposed under the subgroup $U(a+1)$ as
$$
\{T^A \} = \{Y_{IJ}, \ov Y^{I J}, T_I^J \}. \quad\quad\quad I,J = 1,2,\cdots,a+1,
$$
in which $Y_{IJ}=-Y_{JI}$, $(Y_{IJ})^\dagger =\ov Y^{I J}$ and $(T_I^J)^\dagger = T_J^I$.
 They satisfy the Lie-Algebra
\begin{eqnarray}
  [ Y_{IJ} , Y_{KL} ] &=&  0,   \quad\quad\quad h.c., \nonumber \\
\quad [\ov Y^{IJ}, Y_{KL} ] &=& \delta_K^I T_L^J - \delta_L^I T_K^J - 
\delta_K^J T_L^I
  + \delta_L^J T_K^I,  \label{all} \\
\quad [T_I^J, Y_{KL} ] &=& -\delta_K^J Y_{IL} - \delta_L^J Y_{KI}, \quad\quad
h.c., \nonumber \\
\quad  [T_I^J, T_K^L ] &=& -\delta_K^J T_I^L + \delta_I^L T_K^J. \nonumber
\end{eqnarray}
$T_J^I$ are the generators of $U(a+1)$, while $Y_{IJ}$ and $\ov Y^{IJ}$ the broken generators. The Casimir is given by 
$$
{1\over 2}(Y_{IJ}\ov Y^{IJ} +\ov Y^{IJ}Y_{IJ}) + T_I^J T_J^I.
$$
The local coordinates of the coset space $SO(2a+2)/U(a+1)$ are denoted by $\phi_{IJ}$ and $\bar\phi^{IJ}$, correspondingly to the broken generators. Hereinafter upper or lower indices of the coordinates stand for complex conjugation. Therefore lowering or raising  them should be done by writing the metric $g_{IJ}^{\ \ \ KL}$ or $(g^{-1})_{IJ}^{\ \ \ KL}$ explicitly. 

\vspace{1cm}

The Killing vectors $R^A_{\ \ MN}(\phi) $ and $R^{A\ MN}(\bar \phi)$ are respectively non-linear realizations of the Lie-algebra (\ref{all}) on $\phi_{MN}$ and $\bar\phi^{MN}$:
\begin{eqnarray}
R^A_{\ \ MN} \equiv -i[T^A, \phi_{MN}], \quad\quad c.c.. \label{non}
\end{eqnarray}
They are given by
\begin{eqnarray}
R^{IJ}_{\ \ \ MN} &=& i\delta_{MN}^{IJ} (\equiv  \delta_M^I\delta_N^J - \delta_M^J\delta_N^I),  \nonumber \\
R_{IJ\ MN} &=& i(-\phi_{IM}\phi_{JN}+\phi_{IN}\phi_{JM}), \label{a2} \\
R_{I\ MN}^J &=& i(\delta_M^J \phi_{IN} + \delta_N^J \phi_{MI}).\nonumber
\end{eqnarray}
Then the K\"ahler potential  is found according to the formula (28) in \cite{AM}:
$$
 K = {1\over 2}\log\det [1 +  Q],
$$
where $ Q_M^N = \phi_{ML}\bar\phi^{NL}$. 
Indeed by the Lie-variation it transforms as
\begin{eqnarray}
\epsilon^A {\cal L}_{R^A}  K &\equiv& {1\over 2} \epsilon^A (R^A_{\ \ MN}{\partial\over \partial \phi_{MN}} + R^{A\ MN}{\partial\over \partial \bar\phi^{MN}})  K  \nonumber \\
  &=& -{i\over 2}(\bar\epsilon^{MN}\phi_{MN} 
                   + \epsilon_{MN}\bar\phi^{MN}).  \label{a3}
\end{eqnarray}
Note that for the case of $a=2$ the K\"ahler potential takes a simple form such that
\begin{eqnarray}
 K = \log (1 + \phi_{12}\bar \phi^{12} + \phi_{23}\bar \phi^{23} + \phi_{31}\bar \phi^{31}).  \label{ka}
\end{eqnarray}
By using the formula (\ref{form}) with (\ref{a2}) and (\ref{a3})  we obtain the Killing potentials
\begin{eqnarray}
M^{IJ} &=& -[{1\over 1+Q }]^I_L\bar\phi^{LJ}, \quad\quad\quad 
 M_{IJ} = -[{1\over 1+Q }]_I^L\phi_{LJ},  \nonumber \\
 M_I^J &=& -[{Q\over 1+Q}]_I^J +{1\over 2}\delta_I^J. \label{a4}
\end{eqnarray}
One can check that they transform as the adjoint representation of $SO(2a+2)$ by the Lie-variation with respect to the Killing vectors (\ref{a2}) and satisfy
$$
M_{IJ}M^{IJ} + M_I^JM_J^I = {1\over 4}(a+1).
$$

\vspace{1cm}

The K\"ahler coset space $SO(2a+1)/U(a)$ can be constructed 
 from the knowledge of the coset space $SO(2a+2)/U(a+1)$, as has been done for the case of $a=2$ in Section 5. To this end we decompose the generators of $SO(2a+2)$ under $U(a)$ as
$$
 \{T^A\} = \{Y_{ij}, \ov Y^{ij}, Y_{i\ a+1}, \ov Y^{i\ a+1}, T_i^j, T_i^{a+1}, T^i_{a+1}, T_{a+1}^{a+1} \}.
$$
with $i,j=1,2,\cdots,a$. They are grouped in the irreducible  representations of $SO(2a+1)$
as   
\begin{eqnarray}
&\underline {adjoint  \ rep.}: & \left\{
\begin{array}{ll}
   Y_{i\ a+1} - T_i^{a+1} \ (= X_i) , & \quad\quad\quad h.c., \\
  &  \\
     \sqrt 2 Y_{ij} \ (=  X_{ij}), &  \quad\quad\quad  h.c., \\
  &  \\
  \sqrt 2 T_i^j \ (= H_i^j ),  & 
\end{array}\right.   \label{basis}   \\
      & \quad &     \nonumber    \\
&\underline {funda.\  rep.}: & \left\{
\begin{array}{ll}
 Y_{i\ a+1} + T_i^{a+1}, &  \quad\quad\quad h.c., \\
 &  \\
  \sqrt 2 T_{a+1}^{a+1}.
\end{array}\right.   \label{fun}  
\end{eqnarray}
In (\ref{basis})  the generators of $SO(2a+1)$ are given in the basis of $U(a)$. The Casimir of $SO(2a+1)$  takes the form 
$$
X_i\ov X^i + \ov X^i X_i + {1\over 2}(X_{ij}\ov X^{ij} + \ov X^{ij}X_{ij} ) + H_i^j
H_j^i.
$$
$X_i$, $X_{ij}$ and 
their hermite conjugates are the broken generators. The coset space  $SO(2a+1)/U(a)$ is parametrized by the coordinates corresponding to them, denoted respectively by 
$z_i$, $z_{ij}$ and their complex conjugates. 
  To obtain the Killing vectors of $SO(2a+1)/U(a)$ we take from 
  (\ref{a2})  the subset of the Killing vectors which realize the isometry $SO(2a+1)$\begin{eqnarray}
R^i_{\ MN} &=& -i[\ov X^i , \phi_{MN} ],  \quad 
 R_{i\ MN} = -i[X_i , \phi_{MN} ],    \nonumber \\
R^{ij}_{\ MN}  &=&  -i[\ov X^{ij}, \phi_{MN} ],   \quad\quad\quad 
R_{ij\ MN}  =  -i[X_{ij}, \phi_{MN} ], \label{a5} \\
R_{i\ MN}^j &=& -i[H_i^j , \phi_{MN} ].  \nonumber 
\end{eqnarray}
We 
identify the coordinates of $SO(2a+2)/U(a+1)$ with those of $SO(2a+1)/U(a)$ as 
\begin{eqnarray}
 \phi_{i\ a+1} \equiv z_i, \quad\quad \phi_{ij} \equiv \sqrt 2 z_{ij}, \quad\quad c.c..
  \label{id} 
\end{eqnarray}
 Then from (\ref{a5}) we find
\begin{eqnarray}
R^i_{\ m} &=& i\delta^i_m, \quad\quad\quad R_{i\ m} = i(\sqrt 2 z_{im} -z_iz_m ),  
\nonumber \\
R^{ij}_{\ \ m} &=& 0, \quad\quad\quad R_{ij\ m} = 2i(-z_{im}z_j + z_{jm}z_i ), \nonumber \\
R^j_{i\ m} &=& \sqrt 2 i\delta^j_m z_i, \nonumber \\
R^i_{\ mn} &=& {i\over \sqrt 2}(\delta^i_m z_n - \delta^i_n z_m ), \quad\quad\quad 
R_{i\ mn} = i(z_{im}z_n - z_{in}z_m ),  \label{subkil}  \\
R^{ij}_{\ \ mn} &=& i\delta^{ij}_{\ \ mn}, \quad\quad\quad 
 R_{ij\ mn} = 2i(-z_{im}z_{jn} + z_{in}z_{jm}), \nonumber \\
R^j_{i\ mn} &=& \sqrt 2 i(\delta^j_m z_{in} + \delta^j_n z_{mi} ). \nonumber 
\end{eqnarray}
They  realize the Lie-algebra of $SO(2a+1)$  non-linearly on the coordinates $z_{mn}$ and $z_m$, given in the basis (\ref{basis}). Therefore they are the Killing vectors of $SO(2a+1)/U(a)$. 
 The K\"ahler potential is given in the same form as (\ref{ka})
$$
K = {1\over 2} \log\det [1 + Q],
$$
but with the identification (\ref{id}), i.e., 
$$
Q^N_M = \left(
\begin{array}{cc}
 2z_{ml}\bar z^{nl} + z_m\bar z^n &  -\sqrt 2 z_{ml}\bar z^l \\
   & \\
 -\sqrt 2 z_l\bar z^{nl}   & z_l\bar z^l  
\end{array}
\right).
$$
One can check that 
it transforms according to (\ref{a3}) 
by the Lie-variation with respect to  the Killing vectors (\ref{subkil}) of  $SO(2a+1)/U(a)$.  The Killing potentials of $SO(2a+2)/U(a+1)$, given by (\ref{a4}), decomposed   into the two subsets corresponding to (\ref{basis}) and (\ref{fun}). With the identification (\ref{id}) the subset in the adjoint representation gives the Killing potentials of $SO(2a+1)/U(a)$, while the one in the fundamental representation the harmonic functions of $S^{2a}$.

\vspace{3cm}

\hspace{3cm}

\end{document}